\documentclass[aps,prl,twocolumn,amsmath]{revtex4}
\usepackage{graphicx}

\newcommand{\cro}{Cr$_2$O$_4$}
\newcommand{\vo}{V$_2$O$_4$}

\def\ket#1{| #1 \rangle}

\begin{document}

\title{Structural, orbital, and magnetic order in vanadium spinels}
\author{O. Tchernyshyov}
\affiliation{Department of Physics and Astronomy, The Johns Hopkins University,
Baltimore, Maryland, 21218}

\begin{abstract}
Vanadium spinels (Zn\vo, Mg\vo, and Cd\vo) exhibit a sequence of
structural and magnetic phase transitions, reflecting the interplay of
lattice, orbital, and spin degrees of freedom.  We offer a theoretical
model taking into account the relativistic spin-orbit interaction,
collective Jahn-Teller effect, and spin frustration.  Below the
structural transition, vanadium ions exhibit ferroorbital order and
the magnet is best viewed as two sets of antiferromagnetic chains with a
single-ion Ising anisotropy.  Magnetic order, parametrized by two
Ising variables, appears at a tetracritical point.
\end{abstract}

\maketitle

Frustrated magnetic systems have generated interest among researchers
for more than a decade
\cite{Schiffer96,Greedan01,Moessner01}. Frustration refers to the
inability of spins to satisfy conflicting interactions resulting in a
large degeneracy of the ground state at the classical level.  The
effect is particularly strong for spins with antiferromagnetic
interactions on the three-dimensional network of corner-sharing
tetrahedra, known as the pyrochlore lattice \cite{Moessner98}.  Magnets
of this kind have rather unusual properties.  For
example, spinels containing magnetic ions Cr$^{3+}$ (Zn\cro, Mg\cro,
and Cd\cro) undergo a magnetoelastic phase transition
\cite{Lee00,Tch02}, even though they have no spin chains commonly
associated with the spin-Peierls effect \cite{Bray75}.

In this Letter I discuss magnetic and structural properties of another
remarkable group of magnetic spinels containing V$^{3+}$ ions (Zn\vo,
Mg\vo, and Cd\vo).  These compounds represent a step up in complexity
from their chromium counterparts because vanadium ions have an orbital
degree of freedom---{\em in addition to} lattice vibrations and spin.
The interplay of orbital, spin, and vibrational motion against the
backdrop of a high spatial symmetry leads to a sequence of phase
transitions involving changes in structural and magnetic properties.
The physics involves the relativistic spin-orbit interaction,
collective Jahn-Teller effect, and spin frustration.

The vanadium spinels undergo a structural phase transition at a
temperature $T_s$ and a magnetic ordering at a lower temperature
$T_N$.  The structural distortion lowers the crystal symmetry from
cubic $Fd\bar{3}m$ to tetragonal $I4_1/amd$ \cite{Onoda03,Reehuis03},
with a slightly shorter lattice period along the $c$ axis.  The N\'eel
temperature $T_N$ is rather small compared to the Curie-Weiss scale
$\Theta_{\rm CW}$, a sure sign of strong spin frustration.  Ordered
magnetic moments point along the $c$ axis and have the length $\langle
\mu \rangle < \mu_B$ at helium temperatures.  Magnetic susceptibility
is history-dependent both below and above $T_N$
\cite{Ueda97,Reehuis03}.  The numbers for Zn\vo\/ are representative
of the entire family: $c/a = 0.994$, $T_s = 51$ K \cite{Ueda97}, $T_N
= 42$ K \cite{Mamiya97}, $\Theta_{\rm CW} = 420$ K \cite{Mamiya95},
and $\langle \mu \rangle = 0.63 \mu_B$ \cite{Reehuis03}.

The main difference between Cr$^{3+}$ and V$^{3+}$ ions is the number
of electrons in the partially filled $3d$ shell.  The crystal field
splits the $3d$ quintet into a low-energy triplet $t_{2g}$ and a
high-energy doublet $e_g$, separated by a gap of roughly 2 eV.  In
accordance with Hund's rules, the three $d$ electrons of Cr$^{3+}$
have spin $S=3/2$ and occupy all three triplet levels.  The ion has a
spin degree of freedom only.  In contrast, the two $d$ electrons of
V$^{3+}$ have spin $S=1$ and occupy two out of three $t_{2g}$
orbitals.  

\begin{figure}
\includegraphics[width=\columnwidth]{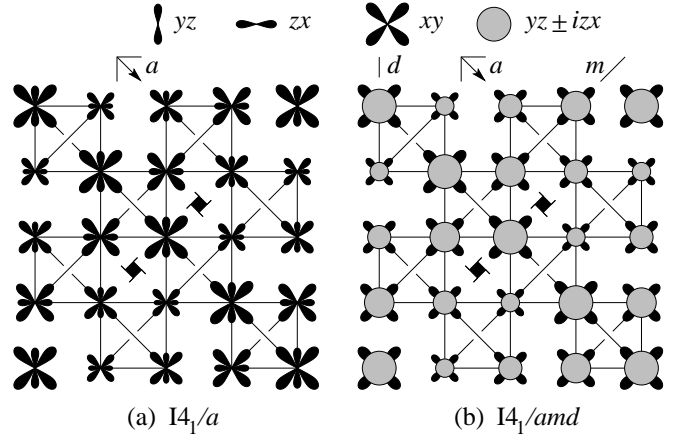}
\caption{Orbital order of vanadium ions below the structural phase
transition.  A view down the [001] direction.  Smaller symbols are
farther away.  (a) The result of Tsunetsugu and Motome
\cite{Tsunetsugu03}.  (b) This work.  }
\label{fig-OO}
\end{figure}

Tsunetsugu and Motome \cite{Tsunetsugu03} have recently developed a
model explaining structural and magnetic properties of the vanadium
spinels.  Following the approach of Kugel and Khomskii, they focus on
Coulomb and exchange interactions between magnetic ions.  However,
their model is at odds with experimental observations: the orbital
order predicted by the theory is incompatible with the spatial
symmetry of the tetragonal phase, identified as $I4_1/amd$
\cite{Onoda03,Reehuis03}.  The staggering of ions with empty $yz$ and
$zx$ orbitals would break reflections $m$ in the planes $(110)$ and
$(1\bar{1}0)$ and diamond glides $d$ in the planes $(100)$ and
$(010)$, lowering the symmetry to $I4_1/a$ [Fig.~\ref{fig-OO}(a)].

I propose an alternative theory that is compatible with the structural
data and predicts rather nontrivial magnetic behavior.  In the new
model, the structural order arises from simple single-ion physics: the
interplay of the JT effect and relativistic spin-orbit interaction.
The $t_{2g}$ orbitals are {\em partially} ordered: their occupation
numbers satisfy $n_{yz} = n_{zx} < n_{xy}$ [Fig.~\ref{fig-OO}(b)].
The distorted magnet is best decribed as a collection of
antiferromagnetic chains, running in the directions $[110]$ and
$[1\bar{1}0]$, with an effective spin $J = 2$ and a single-ion
anisotropy of the Ising type.  Parallel chains are weakly coupled; the
coupling of orthogonal chains is frustrated by symmetry.  As a result,
the N\'eel phase is characterized by two independent Ising order
parameters; the magnetic transition occurs at a {\em tetracritical}
point.  Two nearby magnetic phases exist in which only one half of the
chains order.  Finally, memory effects in magnetic susceptibility are
tied to the spontaneous breaking of the lattice symmetry.

{\em Choice of the model.}  The main difficulty in obtaining a
consistent theoretical picture of the vanadium spinels is a poor
separation of energy scales.  The $t_{2g}$ orbitals of a B site in
spinels point away from the oxygen ligands and thus have a relatively
weak JT coupling \cite{Goodenough63} comparable in magnitude to the
relativistic spin-orbit coupling.  (A recent estimate
\cite{Mizokawa96} puts the JT energy in YVO$_3$ at 20--60 meV.  The
spin-orbit splitting of the $^3F_2$ and $^3F_3$ levels in a free
V$^{3+}$ equals 39 meV \cite{Moore71}.)  Furthermore, the effective
interaction of spin and orbital degrees of freedom on adjacent
vanadium ions has a similar energy scale \cite{Tsunetsugu03}.  It is
not at all obvious which of the three interactions plays the dominant
role.

To make progress, one can adopt a phenomenological approach and
investigate three simple limits, in which the dominant interactions
are, respectively, (I) the JT coupling, (II) the spin-orbit coupling,
(III) the Kugel-Khomskii interaction.  Model III \cite{Tsunetsugu03}
yields orbital order incompatible with the observed symmetry of the
tetragonal phase \cite{Onoda03,Reehuis03}.  We therefore explore
limits I and II.

{\em Basis states and notation.}  If the energy difference between the
$^3P$ and $^3F$ electronic terms of the $3d^2$ configuration ($\approx
1$ eV) is neglected in comparison with its crystal-field splitting in
a VO$_6$ octahedron ($\approx 2$ eV), the $^3T_{1g}$ ground state of a
V$^{3+}$ ion has the $t_{2g}^2 e_g^0$ configuration.  Its orbital part
is triply degenerate and has the following basis:
\begin{eqnarray}
\ket{X} &=& (\ket{zx} \ket{xy} - \ket{xy}\ket{zx})/\sqrt{2},
\nonumber
\\
\ket{Y} &=& (\ket{xy} \ket{yz} - \ket{yz}\ket{xy})/\sqrt{2},
\label{eq-XYZ}
\\
\ket{Z} &=& (\ket{yz} \ket{zx} - \ket{zx}\ket{yz})/\sqrt{2}.
\nonumber
\end{eqnarray}
In this Hilbert space, the operator of orbital momentum ${\bf L}$ 
can be expressed in terms of an effective angular momentum ${\bf L'}$
of length $L'=1$ such that 
\begin{equation}
L'_z \ket{X} = i \ket{Y}, \hskip 5mm
L'_z \ket{Y} = -i \ket{X}, \hskip 5mm
L'_z \ket{Z} = 0 
\end{equation}
etc.  It can be verified that ${\bf L} = -\alpha{\bf L}'$, where
$\alpha = 1$ in the approximation made (\ref{eq-XYZ}).  Inclusion of
electronic correlations mixes in the states $t_{2g}^1 e_g^1$ and
$t_{2g}^0 e_g^2$ and puts $\alpha$ somewhere in the range between 1
and 3/2 \cite{Abragam51}.

{\em Model I.}  A tetragonal distortion of a VO$_6$ octahedron along
the $c$ axis splits the $^3T_{1g}$ state into a doublet and a singlet.
The splitting can be mimicked by the traceless operator 
\begin{equation}
V_{\rm JT} = \delta[{L'}_z^2 - L'(L'+1)/3],
\label{eq-JT}
\end{equation}
where $\delta$ is the tetragonal strain expressed in energy units.
For a $3d^2$ ion, $\delta>0$ for an elongated octahedron
\cite{Goodenough63}.  The lowest-energy state is $\ket{Z}$, in which
the $xy$ orbital is empty and the energy is $-2\delta/3$.  In the
opposite case (a flattened octahedron, $\delta<0$), the orbital ground
state is doubly degenerate and has a higher energy $-|\delta|/3$.
Inclusion of the elastic energy $\mathcal O(\delta^2)$ and
minimization with respect to $\delta$ yields a ground state with
$\delta>0$.  Thus Model I, dominated by the JT coupling, predicts an
elongation of VO$_6$ tetrahedra contrary to the data
\cite{Onoda03,Reehuis03,Ueda97,Mamiya97,Lee03}.

{\em Model II.} The relativistic spin-orbit coupling in V$^{3+}$
\begin{equation}
V_{LS} = \lambda ({\bf L \cdot S}) = -\alpha \lambda ({\bf L' \cdot S})
\end{equation}
has the strength $\lambda \approx +20$ meV \cite{Mizokawa96}.  The
``total angular momentum'' ${\bf J' = L' + S}$ is a conserved
quantity.  The lowest-energy state is the $J'=2$ quintuplet separated
by the energy gap $2 \alpha \lambda$ from the $J'=1$ levels.

Inclusion of the JT interaction (\ref{eq-JT}) as a perturbation
($\delta \ll \lambda$) splits the $J'=2$ quintuplet into three levels.
With the aid of the Wigner-Eckart theorem, one finds
\begin{equation}
V_{\rm JT} = (\delta/2)[{J'}_z^2 - J'(J'+1)/3].
\label{eq-JT-J}
\end{equation}
Remarkably, the ground state has the same energy $-|\delta|$
regardless of the sign of the $\delta$.  In an elongated VO$_6$
octahedron ($\delta>0$), the ground state is nondegenerate and has
$J_z = 0$; its magnetic moment $-\mu_B(-\alpha {\bf L}'+ 2 {\bf S})$
is quenched.  In a flattened octahedron ($\delta<0$), the ground state
is a non-Kramers doublet $J'_z = L'_z + S_z = \pm 2$.  Inclusion of
the elastic energy $\mathcal O(\delta^2)$ fails to determine the sign
of the tetragonal JT distortion.  It will be determined by several
competing perturbations, such as higher-order JT terms, Coulomb and
spin-exchange interactions between vanadium ions.

The effect of Coulomb interaction can be crudely estimated by counting
the number of filled $t_{2g}$ orbitals with the largest overlap.  For
example, the Coulomb energy of a [110] V--V bond is raised if the $xy$
orbitals are occupied in both ions \cite{Tsunetsugu03}.  In a state
with uniform orbital order, the average occupations of the orbitals
are as follows:
\begin{equation}
\begin{array}{lll}
J'_z = \pm 2: 
& \langle n_{yz} \rangle = \langle n_{zx} \rangle = 1/2, 
&\langle n_{xy} \rangle = 1,
\\
J'_z = 0: 
& \langle n_{yz} \rangle = \langle n_{zx} \rangle = 5/6,
&\langle n_{xy} \rangle = 1/3.
\end{array}
\label{eq-n}
\end{equation}
The expectation value of the Coulomb energy, summed over
nearest-neighbor V--V pairs, will be proportional to $\langle n_{yz}
\rangle^2 + \langle n_{zx} \rangle^2 + \langle n_{xy} \rangle^2$.
This expression evaluates to 3/2 for both $J'_z=0$ and $J'_z = \pm2$
ground states.  Thus the Coulomb repulsion between $V^{3+}$ ions is
the same in the elongated and flattened ground states of VO$_6$
octahedra and does not influence the outcome of their competition.

(The accidental degeneracy of the Coulomb term can be traced to the
degeneracy of the flattened and elongated JT states for $J' = 2$.
Both the Coulomb and JT energies can be expressed in terms 
of the operators ${J'}_z^2-J'(J'+1)/3$ and ${J'}_x^2 -
{J'}_y^2$.  It is therefore likely that the conclusion reached in the
previous paragraph is not sensitive to the approximations made in this
paper.)

Exchange energy, on the other hand, clearly favors the magnetic ground
states $J'_z = \pm2$ because it can be made negative by an appropriate
choice of spin orientations.  In contrast, exchange energy is zero if
the ions are in the nonmagnetic state $J'_z = 0$.  Anharmonic terms of
the JT energy $\mathcal O(\delta^3)$ can favor either sign of
$\delta$.  Therefore, one cannot deduce the sign of the distortion on
general theoretical grounds.  Nonetheless, experimental observation of
antiferromagnetic order in Zn\vo\/ at low temperatures points to the
magnetically active states $J'_z = \pm2$ characterized by a flattening
of the VO$_6$ octahedra along the $c$ axis.  For a given direction of
the spin, $S_z = \pm 1$, one of the electrons occupies the $xy$
orbital, whereas the other is in the state $\ket{yz} \pm i\ket{zx}$,
whose actual orbital moment $L_z = -\alpha L'_z = \mp \alpha$ points
opposite to the spin.  Equal occupation of the $yz$ and $zx$ orbitals
in such a state is in agreement with the observed spatial symmetry of
the crystal ($I4_1/amd$).  Magnetic moments of length
$(2-\alpha)\mu_B$, where $1\leq \alpha < 3/2$, point along the $c$
axis, as found experimentally.  Thus phenomenological Model II, with
strong relativistic spin-orbit coupling, a moderate JT distortion, and
weak V--V interactions, is compatible with experimental observations.

{\em Structural order and symmetry-breaking fields.}  Below $T_s$ the
VO$_6$ octahedra flatten along one of the three major axes.  From the
symmetry viewpoint, the long-range order is similar to that of the
3-state Potts ferromagnet.  A two-component order parameter ${\bf f} =
(f_1, f_2)$ can be constructed, e.g. from the lattice constants:
\begin{subequations}
\begin{equation}
f_1 = (a+b-2c)/\sqrt{6},
\ \ 
f_2 = (a-b)/\sqrt{2}.
\label{eq-OPa}
\end{equation}
Other observables can be used to define the order parameter.  The
lattice constants $(a,b,c)$ can be replaced with orbital populations
$(\langle n_{yz} \rangle, \langle n_{zx} \rangle, \langle n_{xy}
\rangle)$, or with magnetic susceptibilities $\chi_{ij} = \partial M_j
/ \partial H_i$:
\begin{equation}
f_1 = (\chi_{xx} + \chi_{yy} - 2 \chi_{zz})/\sqrt{6}, 
\ \ 
f_2 = (\chi_{xx} - \chi_{yy})/\sqrt{2}. 
\label{eq-OPb}
\end{equation}
\end{subequations}

Explicit construction of the order parameter allows us to identify
symmetry-breaking fields that can be applied to select a particular
phase with broken symmetry.  The most obvious among these is the
two-component (tetragonal and orthorombic) stress ${\bf h} = (h_1,
h_2)$:
\begin{subequations}
\begin{equation}
h_1 = (e_{xx} + e_{yy} - 2e_{zz})/\sqrt{6}, 
\ \ 
h_2 = (e_{xx}-e_{yy})/\sqrt{2}.
\label{eq-SBFa}
\end{equation}
Cooling under a weak uniaxial stress applied along the $c$ axis will
produce an ordered phase with $a=b>c$.

A less obvious symmetry-breaking perturbation is magnetic field ${\bf
H}$.  The magnetic energy density of a sample can be written as
$-(\chi_{xx} H_x^2 + \chi_{yy} H_y^2 + \chi_{zz} H_z^2)/2.$ Unless the
field is parallel to a $\langle 111 \rangle$ direction, it lowers the
energies of different ordered phases by different amounts.  The
symmetry-breaking field is
\begin{equation}
h_1 = (H_x^2 + H_y^2 - 2 H_z^2)/\sqrt{6}, 
\ \ 
h_2 = (H_x^2 - H_y^2)/\sqrt{2}. 
\label{eq-SBFb}
\end{equation}
\end{subequations}

These considerations could explain memory effects in magnetic
susceptibility of Zn\vo\/ above the N\'eel temperature $T_N$
\cite{Ueda97,Reehuis03}.  A crystal cooled through the structural
transition in magnetic field chooses the distorted phase with the
lowest magnetic energy.  The difference between $\chi_{\rm FC}$ and
$\chi_{\rm ZFC}$ in powder samples may simply reflect the presence of
the order parameter (\ref{eq-OPb}) below $T_s$ and its sensitivity to
the symmetry-breaking field (\ref{eq-SBFb}).  It would be highly
desirable to study these memory effects in single crystals.

\begin{figure}
\includegraphics[width=\columnwidth]{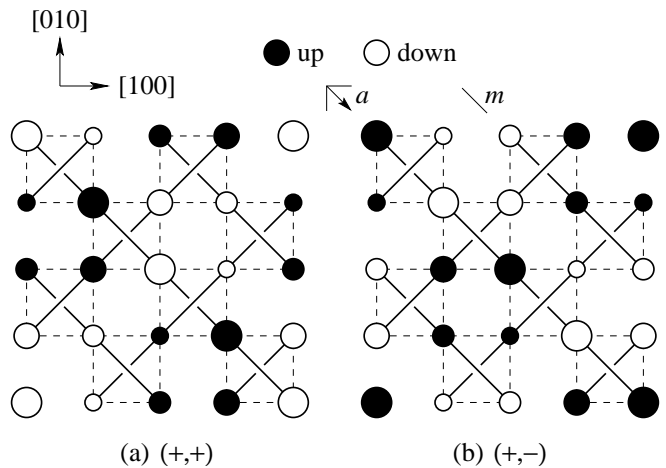}
\caption{Two of the four magnetically ordered states.  The signs 
refer to the two Ising order parameters
$(\sigma_{[110]},\sigma_{[1\bar{1}0]})$.  Ground states (a) and (b)
are related by a (110) reflection $m$ or by a (110) glide
$a$ in a $(001)$ plane.}
\label{fig-Neel}
\end{figure}

{\em Aniferromagnetic Ising chains with frustrated coupling.}  In the
distorted state, the magnetic system, comprising spins and orbital
moments, can be thought of as a collection of chains running in the
$[110]$ and $[1\bar{1}0]$ directions (Fig.~\ref{fig-Neel}).  Because
the $xy$ orbitals are always occupied, the largest exchange coupling
takes place along the chains.  Exchange coupling along the other
$\langle 110 \rangle$ directions is expected to be substantially
weaker because the relevant orbitals ($yz$ and $zx$) are only
half-filled.  Onoda and Hasegawa \cite{Onoda03} have extracted the
values of $J_{[110]} \approx 5$ meV and $J_{[011]} = J_{[101]} \approx
3$ meV for Cd\vo.  In addition to the single-ion Ising anisotropy
(\ref{eq-JT-J}), which is a product of orbital order and spin-orbit
coupling, the exchange interactions are likely anisotropic as well.

As the distorted magnet is cooled further, spin correlations develop
first along the chains, as indeed observed \cite{Lee03}.  If the
interchain coupling is weak, long-range magnetic order will set in at
a low temperature, when each chain already has a large correlation
length.  In that case, the magnetic transition will have almost no
effect on the energetics of the spins.  Indeed, measurements of
specific heat in Mg\vo\/ reveal a very weak singularity at the
magnetic transition \cite{Mamiya97}.

While magnetic interactions between parallel chains are expected to be
weak because they are well separated in space, the coupling between
crossing chains may be substantial: four of their spins are nearest
neighbors forming a tetrahedron (Fig.~\ref{fig-Neel}).  However, in
this case the coupling is frustrated by the lattice symmetry: a
(110) reflection $m$ (or a glide $a$) changes the sign of staggered
magnetization on the chain running in the direction $[1\bar{1}0]$ but
preserves staggered magnetization on the [110] chain.  Therefore, even
if the crossing-chain coupling is numerically comparable to the
intrachain coupling, it is still rendered weak by geometrical
frustration.

{\em Magnetic order.}  Neutron scattering indicates that adjacent
parallel chains have equal values of staggered magnetization
(Fig.~\ref{fig-Neel}).  Because of the geometrical frustration, the
system has {\em four} ground states: those shown in
Fig.~\ref{fig-Neel} and their time-reversed copies.  These ground
states can be characterized by two independent Ising order parameters:
one staggered magnetization for each set of chains, $\sigma_{[110]}$
and $\sigma_{[1\bar{1}0]}$ \cite{chains-Tsu}.

\begin{figure}
\includegraphics[width=\columnwidth]{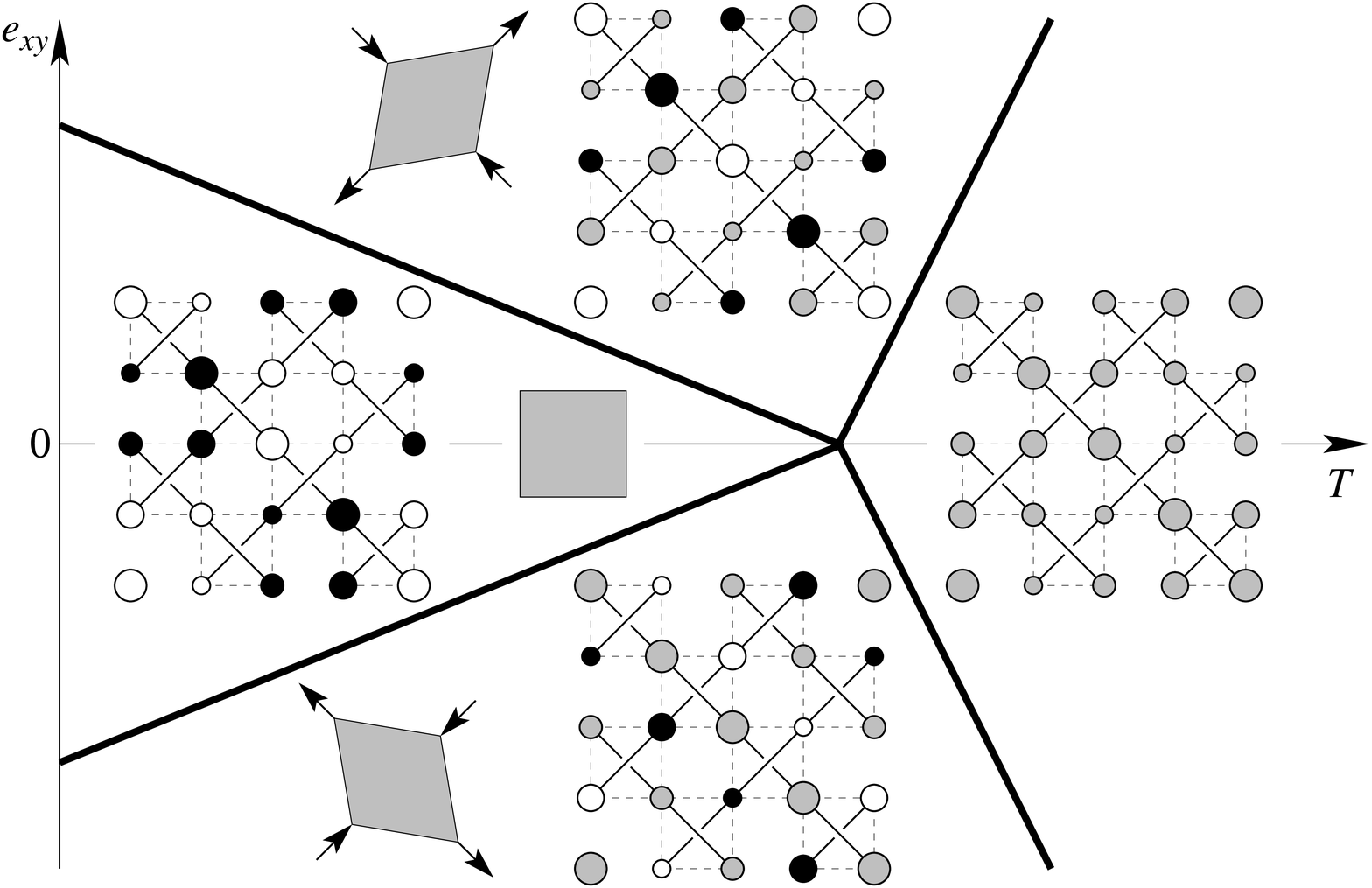}
\caption{The temperature--strain phase diagram in the vicinity of the
tetracritical point $T=T_N$, $e_{xy}=0$.  Gray symbols correspond to
disordered spins.  Thick solid lines are phase boundaries.  Other
notations are same as in Fig.~\ref{fig-Neel}.}
\label{fig-phases}
\end{figure}

{\em Magnetic tetracritical point.}  Both sets of chains appear to be
ordered at the lowest temperatures \cite{Reehuis03}.  In the simplest
scenario, both Ising order parameters develop continuously at the
magnetic transition temperature $T_N$, which is characteristic of a
tetracritical point \cite{Fisher74,Rohrer77}.  One should be able to
observe two additional phases in the vicinity of the tetracritical
point, in which only one set of chains is ordered, while the other is
not.  This may be achieved by creating an orthorombic strain $e_{xy}$
in the vicinity of the magnetic transition.  Doing so will violate the
equivalence of the [110] and $[1\bar{1}0]$ chains while still
preserving the geometrical frustration.  A temperature scan under a
uniaxial [110] stress should reveal two separate magnetic transitions
for each set of chains.  The temperature--strain phase diagram in the
vicinity of the tetracritical point is shown in Fig.~\ref{fig-phases}.

{\em Outlook.} A phenomenological approach taken in this paper is
justified by the complexity of vanadium spinels: the relevant
degrees of freedom include lattice vibrations, orbital motion, and
spin.  It is desirable to put these considerations on a quantitative
footing with the aid of a microscopic theory.  Nonetheless, the
qualitative predictions of this work---a spin gap due to single-ion
anisotropy, a magnetic tetracritical point, and a potential relation
between magnetic memory effects and broken lattice symmetry---can be
tested experimentally.

Upon completion of this work I learned that a similar model had been
considered by D.I. Khomskii.

{\em Acknowledgment}.  I thank C. Broholm, D.I. Khomskii, A. Krimmel,
M.Z. Hasan, S.-H. Lee, D. Louca, M. Onoda, P.G. Radaelli,
O.A. Starykh, and H. Ueda for useful discussions.  This work was
supported in part by the U.S. National Science Foundation under the
Grant No. DMR-0348679.

\end{document}